\documentclass[a4,10pt]{article}
\usepackage[utf8]{inputenc}
\usepackage{mathptmx}
\usepackage[affil-it]{authblk} 
\usepackage[a4paper, margin=1in]{geometry}
\usepackage{sectsty}
\sectionfont{\fontsize{10}{10}\selectfont}
\subsectionfont{\fontsize{10}{10}\selectfont}
\usepackage{titlesec}
\titleformat*{\subsection}{\normalsize\itshape}
\usepackage{indentfirst}
\usepackage{array,multirow}
\usepackage{fancyhdr}
\usepackage{graphicx}
\usepackage{caption}
\usepackage{booktabs}
\usepackage{multirow}
\usepackage{diagbox}
\newcommand{\tabincell}[2]{\begin{tabular}{@{}#1@{}}#2\end{tabular}} 
\usepackage{txfonts}

\title{\huge \textbf{Driving behavior model considering driver's over-trust in driving automation system}}

\author{
        \large \textbf{Hailong LIU~$^{*}$, Toshihiro HIRAOKA~$^{*}$}\\  
        $^{*}$~Nagoya University\\
               \ \\
        \normalsize
        Furo-cho, Chikusa-ku, Nagoya, Aichi, 464-8601, JAPAN\\
        Phone: (+81) 52-747-6977 \\
        E-mail: liu.hailong@mirai.nagoya-u.ac.jp \\
        Co-author’s E-mail: toshihiro.hiraoka@mirai.nagoya-u.ac.jp
           }

\date{}
\begin{document}
\maketitle

\begin{abstract}
Levels one to three of driving automation systems~(DAS) are spreading fast.
However, as the DAS functions become more and more sophisticated, not only the driver's driving skills will reduce, but also the problem of over-trust will become serious.
If a driver has over-trust in the DAS, he/she will become not aware of hazards in time.
To prevent the driver's over-trust in the DAS, this paper discusses the followings:
1) the definition of over-trust in the DAS,
2) a hypothesis of occurrence condition and occurrence process of over-trust in the DAS, and
3) a driving behavior model based on the trust in the DAS, the risk homeostasis theory, and the over-trust prevention human-machine interface.
\end{abstract}

\section{Introduction}
Driving automation system~(DAS) is a dynamically control system of a vehicle. 
It is developed by imitating the driving process of a human driver considering interaction occurring among the driver-vehicle-environment.
In general, a driver recognizes the vehicle state, e.g., velocity, acceleration, and the driving environment, e.g., the location, the shape of road, objects, and hazards, via body sensory information such as visual and auditory information.
After that, the driver takes the recognized the vehicle state and the driving environment into account to decide a driving plan and then drives the vehicle.
As for the DAS, the driving process is the same as the driving process of the human driver.
The DAS equips sensors to replace the driver's three semicircular canals, eyes, and ears to identify the vehicle state and the traffic conditions~\cite{sun2006road,weber2016deeptlr,schneider2016semantic},
e.g., a camera, G-sensors, gyros, a LiDAR, and a microphone.
Then, the DAS controls the vehicle based on analyzing the information observed by those sensors. 
The above process is common to all DAS basically.

Different DASs have different requirements for usage conditions, e.g., driving task and operational domain.
Therefore, {\it SAE international} defined the DASs into five levels~\cite{SAE_j3016_2016} as shown in Table~\ref{SAE}.
Levels one to three of the DASs require the driver to be fully or partially involved in the driving tasks, e.g., part of the vehicle control and emergency intervention.
That is, the driver must not only ``properly monitor'' the driving environment and the state of the DAS all the time while driving, but also need to ``properly handles'' driving task when the situation exceeds the functional limitation of the DAS or when the DAS fails.
Due to the above constraints, it is easy to imagine that the following problems will occur more frequently as the system performance improves:
\begin{description}
\item[1)]~The driver's driving skill~\cite{young2007s} and  situation awareness~\cite{stanton2005driver} will decline as the driving opportunities decreases.
\item[2)]~The prolonged monitoring control for the DAS is extremely likely to increase the mental and physical workload on the driver in the case when the DAS does not (or less) requires driving operation of the driver~\cite{Ohtani201620164429}.
\item[3)]~As the DAS functionality becomes more diverse and more reliable~(reliable does not mean that no errors will occur), the driver may easily over-trust in the DAS when the driver does not clearly understand conditions of use and the functional boundaries of the DAS~\cite{parasuraman1997humans,Uno2010,Carsten2018}.
\end{description}

\begin{table*}[bt]
\captionsetup{justification=centering}
\caption{Five levels of driving automation systems defined by SAE J3016 (2016)~\protect\cite{SAE_j3016_2016}}
\centering
\scalebox{0.9}[1]{ 
\begin{tabular}{@{}cccccc@{}}
\toprule
\multirow{2}{*}{Level} & \multirow{2}{*}{Name} & \multicolumn{2}{c}{Dynamic driving task~(DDT)} & \multirow{2}{*}{\begin{tabular}[c]{@{}c@{}}DDT fallback\end{tabular}} & \multirow{2}{*}{Operational domain} \\ \cmidrule(lr){3-4}
                     &                     & Vehicle control     &  \tabincell{c}{Object and event\\detection and response  }   &                                                                               &                       \\ \midrule
1	& Driver assistance                &{\bf System and driver}     & {\bf Driver}         & {\bf Driver}                                                                       & Limited                   \\
2	& Partial automation             & System         & {\bf Driver}     &{\bf Driver}                                                                           & Limited                   \\
3	& Conditional automation            & System         & System         & {\bf Ready Driver}          & Limited                  \\ 
4	& High automation
& System         & System         & System          & Limited                  \\
5	& Full automation            & System         & System         & System          & No limited                  \\
\bottomrule
\end{tabular}
}
\label{SAE}
\end{table*}

The over-trust in the DAS will put the driver in a danger when the situation is beyond what the DAS can respond.
For example, a fatal accident happened when a driver drives a car with a level two DAS in Jun 2016 at the US.
According to the accident investigation report, there is a high possibility that the driver did not fulfill his duty to observe the driving environment and do the driving task when a large white truck with a trailer crossing in the highway in front was not detected by the camera of the DAS under the influence of strong sunlight.
In the 37 minutes driving before the accident, the driver held the steering wheel only for 25 seconds~\cite{ntsb1}.
This investigation implies that the driver had over-trust in the performance of the level two DAS at first.
The driver may mistakenly believe that the DAS will work safely even hands off during driving and the DAS's environment recognition system can perfectly recognize anything under any condition.
Therefore, the driver did a behavior of over-reliance on DAS -- driving without holding the steering wheel in most of the time.
Meanwhile, the over-confidence in the DAS's ability of environmental understanding might be one of the reasons why the accident happened.
For example, the driver may believe that he/she has understood the DAS, but he/she does not fully understand it in fact.
The worse-case scenario is that the driver may believe that he/she is able to control the vehicle to avoid a danger in time, even if he/she does not hold the steering wheel while using the DAS.
This phenomenon of over-confidence can be explained by Dunning--Kruger effect~\cite{kramer1985application}, which argues that people are difficult to objectively evaluate their ability and skill because they often have a cognitive bias of illusory superiority.

In general, whether the driver has over-trust in the DAS or himself/herself~(over-confidence),
mainly because the driver cannot exactly understand the system design, purpose, mechanism, and ability of the DAS.
An effective approach to solve this problem is to provide opportunities of education and training to the users.
It is especially important to train drivers to deal with a take-over request of the DAS when it cannot maintain the driving task~\cite{Daniele2019}.
Moreover, legal supervision is also important, such as setting up a licensing system of the DAS, like the driving license~\cite{ntsb2}.
This license is issued when the driver is familiar with the DAS and can use it proficiently.
However, each vendor needs to develop the DAS based on the agreed architecture.
This is difficult to be done at the moment. Therefore, information provision by using some HMIs can be considered as an effective way to solve the problem that drivers have over-trust especially in the levels one to three of the DAS.

The final goal of this study is to design a human-machine interface~(HMI) that prevents over-trust while driving a vehicle with the DAS. The levels one to three of the DAS require the driver's participation in completing the driving tasks.
As the first step of our study, this paper discusses the following issues with respect to over-trust:
\begin{description}
\item[1)]~~definition of over-trust in the DAS;
\item[2)]~~the occurrence condition and process of over-trust;
\item[3)]~~proposal of a driving behavior model with an over-trust prevention HMI.
\end{description}

\section{What is over-trust in the DAS?}
This section answers a question ``what is trust?'' before response to a question ``what is over-trust?''. 
It discusses the relationship between trust and reliance, 
and therefore it defines the driver's over-trust in the DAS.

\subsection{Trust and distrust}
Conceptualizing trust and distrust is a big challenge~\cite{mcknight2001trust}, and they are defined and interpreted differently in different fields.
In the psychology field, trust is often conceptualized as a belief, expectancy, or feeling~\cite{rotter1967new}.
It is regarded as an integral feature of human relations~\cite{rempel1985trust,larzelere1980dyadic}.
In management, the words, behaviors, attitudes, and decisions of human or party affect the trust of others in him/her/it, such as competent, open, concerned, and reliable~\cite{culbert1986politics,Mishra1990,gambetta2000can}.

The above definitions or descriptions of trust are abstract.
Trust needs more specific and nuanced explanations.
Lee et al. proposed four dimensions of trust in a human-machine system~\cite{lee1992trust}, such as:
\begin{description}
\item[1)]~~trust in the {\bf foundation} of a system: the system complies with the laws of nature and the orders of society;
\item[2)]~~trust in the {\bf purpose} of a system: the purpose and motivation of a system can be accepted by a user;
\item[3)]~~trust in the {\bf process} of a system:  process or algorithm of the system can be clearly understood;
\item[4)]~~trust in the {\bf performance} of a system: a stable and desirable performance of the system can be expected.
\end{description}
Note that this system will be distrusted even if only one of these dimensions cannot be satisfied.

\subsection{Trust and reliance}
Trust and reliance are fundamentally different, although they are often lumped together.
Giffin explained the reliance as a behavior of trust, especially in dangerous situations~\cite{giffin1967contribution}.
Rotter considered that the behaviors of reliance, such as verbal and written commitments, could be used to express trust~\cite{rotter1980interpersonal}.
Based on the above references, this paper considers trust as a psychological activity, and reliance as a behavioral manifestation of trust.

\subsection{Over-trust in DAS}
\begin{figure}[tb]
\begin{center}
\includegraphics[width=0.5\columnwidth]{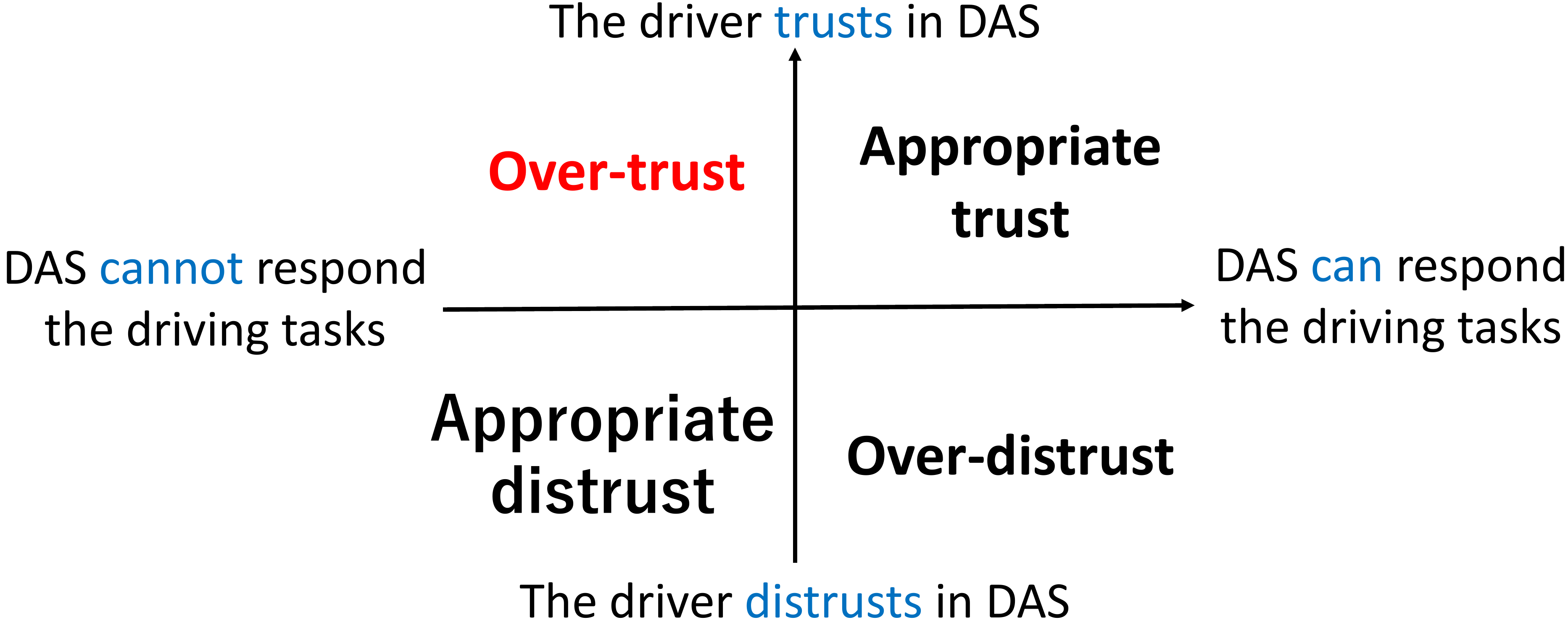}
\end{center}
\captionsetup{justification=centering}
\caption{Two conditions of over-trust in DAS.}
\label{fig:overtrust}
\end{figure}

Inagaki defined over-trust as an incorrect situation of diagnostic decision that the object is trustworthy when it actually is not trustworthy~\cite{inagaki2013human}. 
The over-trust in the DAS is that the DAS cannot respond to driving tasks but the driver trusts it can.
Consequently, there are two judgment conditions of over-trust which are shown in Fig.~\ref{fig:overtrust}:
\begin{description}
\item[1)]~the driver is trusting in the DAS, 
\item[2)]~the DAS cannot respond to driving tasks.
\end{description}
Based on the two conditions of over-trust and four dimensions of trust, there are four types of over-trust: over-trust in foundation, purpose, process, and performance. 
Therefore, the user will become over-trust when the user trusts in the system and the system cannot respond to driving tasks.
One reason for over-trust is that the user has misconceptions about the four dimensions of the system.
The final goal of this study is to prevent over-trust by reducing those misconceptions.

\section{Examples of over-trust in DAS}
The previous section explained the definition of over-trust.
From the perspective of human factor, 
this section describes four types of over-trust based on four dimensions of trust.

\subsection{Over-trust in the foundation of DAS}
The system designer designs the system based on common sense and traffic rules.
However, the traffic rules may differ from country to country.
For example, drivers in some countries such as Japan keep to the left side of the road, while many countries adopt the right-hand traffic. 
If a vehicle equipped with a non-domestic DAS runs in Japan,
it may violate Japanese traffic rules.
There is a risk that an accident may occur due to the misidentification of a traffic sign unique to Japan.
If the driver does not know the difference between traffic rules of Japan and foreign countries,
over-trust in the foundation of the DAS may occur when the driver uses a non-domestic DAS.

\subsection{Over-trust in the purpose of DAS}
The driver may confuse with purposes of different DASs even they have similar driving tasks.
For example, an adaptive cruise control system~(ACC) and an automatic emergency braking system~(AEB) are similar in controlling the longitudinal acceleration of the vehicle. 
However, their purposes are completely different.
The purpose of the ACC is to reduce the physical workload of the driver.
Meanwhile, the purpose of the AEB is to alert a driver to an imminent crash and help them use the maximum braking capacity of the vehicle if the situation becomes critical and the human driver does not respond to the situation.
The ACC does not include the function of the AEB to slow down the vehicle to stop in a dangerous situation although the ACC can automatically accelerate and decelerate a vehicle, because some of the ACC will automatically release control when the speed becomes lower than the lower-limit speed of the ACC.
If the driver trust the ACC can handle the situation which actually requires an emergency brake, then the over-trust in the purpose of the ACC occurs.

\subsection{Over-trust in the process of DAS}
The lane keeping assist system~(LKAS) is a DAS that can alert the driver when the vehicle begins to move out of its lane on freeways and arterial roads by recognizing lane lines via cameras.
Some advanced LKAS cannot only alert the driver, but also automatically control the vehicle's lateral direction to keep it in the lane.
If a driver considers that the LKAS can be used on a road with unclear lane lines, then he/she may has an over-trust in the process of LKAS.
The driver may know nothing about LKAS which needs to detect the lane lines via cameras.

\subsection{Over-trust in the performance of DAS}
The over-trust in the performance of the DAS easily occurs because a driver may be difficult to understand the functional limitation of the DAS.
For example, most AEBs activate the brake system based on using a millimeter-wave radar or a front camera to detect the obstacles.
However, millimeter-wave radar based AEB may cause false detection of obstacles in the rain or snow because raindrops and snowflakes reflect millimeter waves.
Also, the obstacle detection of the camera-based AEB will be disturbed by glare and shadow.
Therefore, the over-trust in the performance of the AEB may occur if the driver does not understand the functional limitations~(performance) of the AEB's.

\section{The judgment of the over-trust}
\subsection{Problem 1: Non-real-time judgment}
When the driver drives the vehicle with the DAS and he/she even knows whether he/she trusts in the DAS or not, it is hard for him/her to tell whether the driver is doing over-trust in the DAS or not.
The reason why such situation happens is that it is hard for the driver to tell (detect) whether the DAS can respond to the driving task or not in real-time by himself/herself.

\subsection{Problem 2: Un-observable property of over-trust}
Both of the two judgment conditions of over-trust include two factors: the driver and the DAS.
The driver cannot fully know the performance of the DAS and the situations where the DAS cannot responded. 
The DAS cannot measure the driver's trust state directly because the trust state is an internal psychological activity.
Therefore, neither the driver nor the monitoring system cannot observe the over-trust directly.

There are two countermeasures to solve the above-mentioned problems;
1) a driver monitoring system to estimate the driver's trust
state, and 
2) a system to predict whether the DAS can respond to the driving situation or not.
Combined with these two systems, over-trust will be predicted before a dangerous event.

\subsection{Who can judge the driver's over-trusts in the DAS?}
The previous subsection explained that the observation of over-trust is impossible because the driver and the system cannot directly measure the state each other.
However, over-trust can be estimated and judged indirectly.
Here, let us consider the method to estimate the over-trust. The driver can spontaneously create a mental model that is used to recognize and to interpret the system by repeatedly using it~\cite{staggers1993mental}.
Therefore, the driver will be able to estimate if the DAS can handle the driving task by comparing the recognized state of the DAS with the predicted state via the mental model.
It is possible for the driver to judge whether he/she over-trusts in the DAS or not by combining the estimated result with his/her trust state.
However, this judgment is usually an afterthought of the driver.
Meanwhile, the driver can use the mental model to predict the failure state of the DAS in advance.
In this case, the trust in the DAS of the driver will decrease and therefore the driver is difficult to over-trust in the DAS.
\begin{figure*}[tb]
\begin{center}
\includegraphics[width=1\linewidth]{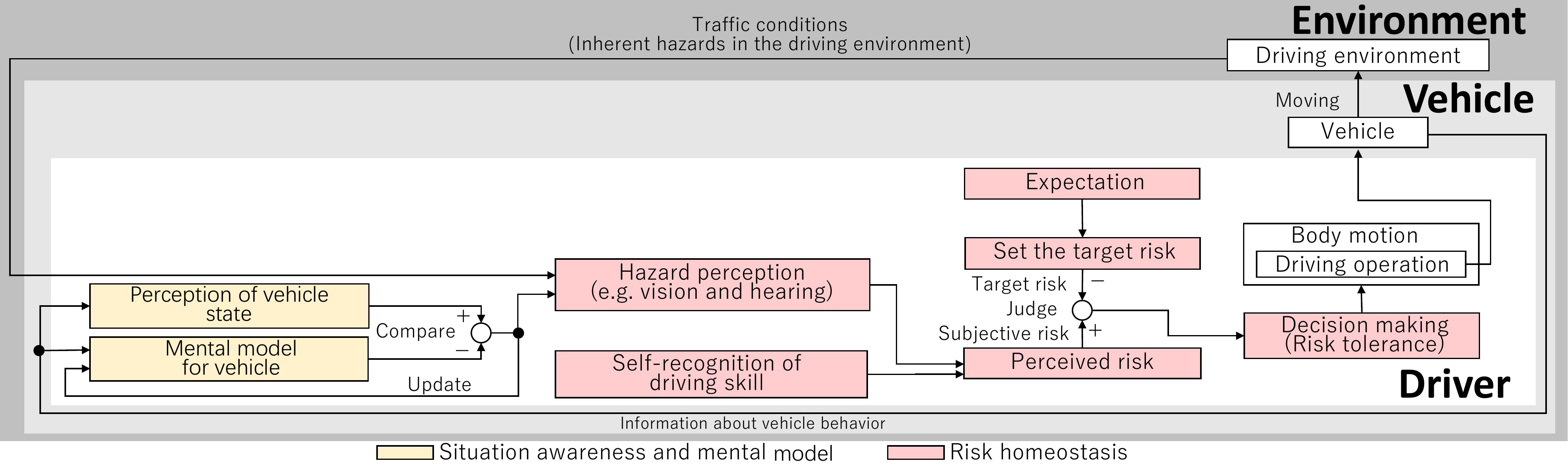}
\end{center}
\captionsetup{justification=centering}
\caption{Driving behavior model considering the mental model and the risk compensation behavior.\\
(Case 1: when driving conventional vehicle)}
\label{fig:BL1}
\end{figure*}

On the other side, 
the system may estimate the driver's preparation state for the driving task based on the driver's biological information and body motions, although the system cannot observe the driver's trust states. 
For example, a driver drives a vehicle with an earlier version of the ACC that cannot stop the vehicle.
In this case, if a system can observe the position of driver's foot by using the camera from the moment when a preceding vehicle starts to decelerate until the moment when the velocity of the ego vehicle drops to the lower limit of the ACC, it may be able to determine whether the driver has over trust in the ACC or not by inferring the driver's readiness for braking operations.

Generally, the answer of {\it ``who can judge the driver's over-trusts in DAS?''} is the ``third-party'' who can know the trust state of the driver and the functional limitation of the DAS. 
There are two types of the ``third-party'' such as
1) a driver who can look back on his own trust state and the operation state of the DAS in the past, and 
2) a monitoring system that can estimate the driver's trust state and the operation state of the DAS by monitoring the driver and the DAS.

\section{Driving behavior model for manual driving}
The DAS, especially levels one to three, are developed on the basis of manual driving vehicles.
Therefore, the study proposes a driving behavior model for manual driving with reference to mental model and risk compensation behavior, as shown in Fig.~\ref{fig:BL1}.
This model includes four parts for driving behavior generation of the driver: 
1) perception of vehicle state,
2) mental models for the vehicle,
3) hazard perception, and
4) risk compensation behavior.

\subsection{Perception of vehicle state}
In the case of manual driving, the driver makes the decision of driving behaviors by considering the situation awareness result of the vehicle state and the perceived hazard in the driving environment.
The vehicle state can be confirmed from the body sensory information of the driver during manual driving.
For example, the speed meter and the tachometer are used to check the information about the vehicle speed and the engine speed by a vision of the driver.
In addition, the driver may be able to sense engine sound for estimating the engine's states, e.g. engine speed, engine trouble.
The driver can feel the acceleration of the vehicle through the inner ear.

\subsection{Mental model for the vehicle}\label{subsection: mental model}
After the driver recognizes the vehicle state by situation awareness, the mental model is used to predict what the vehicle will be operated in such situation.
The mental model~\cite{staggers1993mental} for the vehicle is an explanation of the thought process of the driver about how to operate the vehicle and move it in the real world.
It is based on the long term driving experience of the driver by estimating the relationship between the operation (system input) and the change of the vehicle state (the output of the system).
In simple terms, the mental model can be considered as a prediction model~\cite{richardson1994foundations}.
It also can be seen as a cognitive model for vehicles formed in the driver's brain.
``{\it Stepping on the gas pedal leads to vehicle accelerating}'' is the most simple example of a mental model that is generated from driving experience.
The mental model can be updated by comparing the predicted vehicle state via the mental model and the recognized vehicle state.
Moreover, the comparing result also affects the hazard perception.

\subsection{Hazard perception in manual driving}
The driver tries to understand the vehicle states such as velocity, acceleration, and the traffic environment while driving a vehicle.
The driver perceives the hazards through the vehicle by comparing the recognized situation of the vehicle with the mental model~\cite{grosser2012mental} for the vehicle.
The driver also perceives the hazards through 1) the driving environment by considering traffic conditions, and 2) the recognized vehicle state.
This study assumes that the hazards can be represented by a quantitative value such as the time-to-collision~(TTC) or other indices of the potential collision risk.

\begin{figure*}[tb]
\begin{center}
\includegraphics[width=1\linewidth]{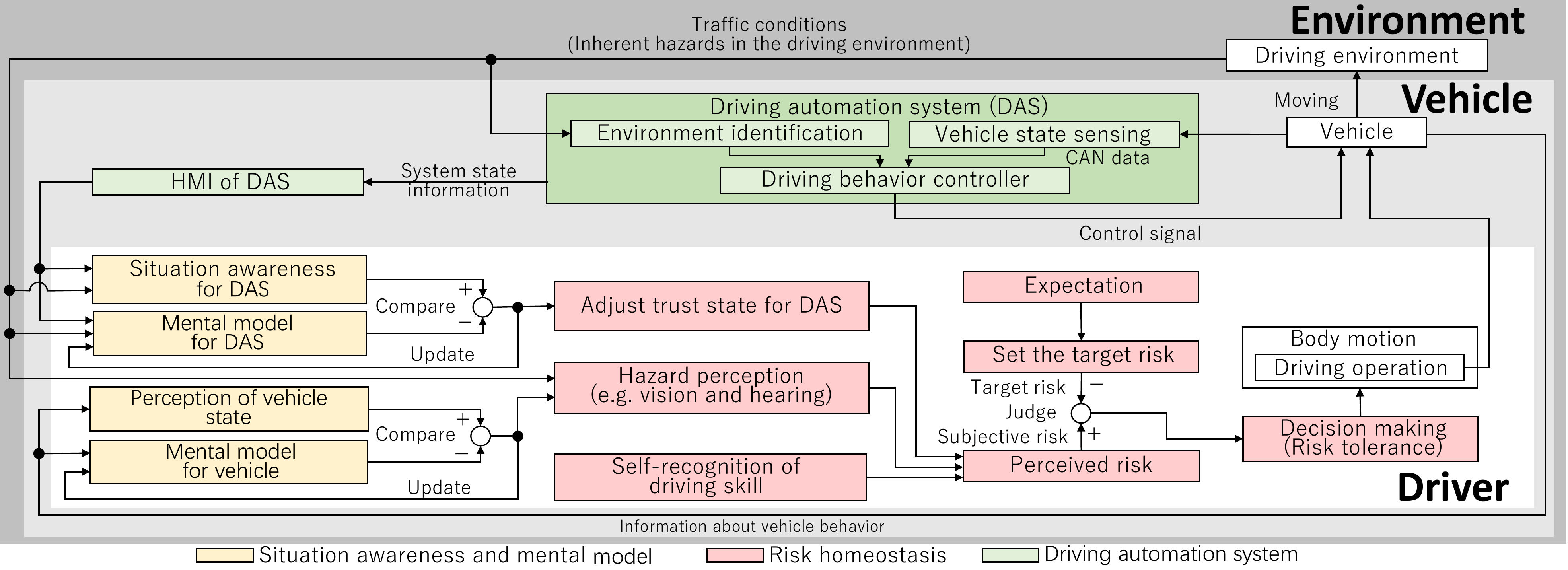}
\end{center}
\captionsetup{justification=centering}
\caption{Driving behavior model considering the mental model, the driver's trust in DAS, and the risk compensation behavior.
(Case 2: when driving vehicle with DAS)}
\label{fig:BL2}
\end{figure*}

\subsection{Risk homeostasis in manual driving}
In manual driving, the driver needs to comprehensively analyze various factors to perceive the risk of the current driving state.
These factors include the perceived hazards and the self-recognized driving skills by the driver.
For example, an experienced driver might think that a complex driving environment with various hazards can be handled easily because he/she has high confidence in driving skills.
The self-recognized driving skills and the perceived hazards jointly affect the risk perceived by the driver.
It should be noted that the wrong self-recognized driving skills can lead to over-trust in oneself.
The authors believe that the occurrence of such over-trust in one's own ability can be prevented through reasonable education and training, although the over-trust in one's own ability is beyond the scope of this manuscript.

According to the risk homeostasis theory~\cite{wilde1982theory},
the driver decides his/her driving behavior (decision making) by comparing perceived risk with the acceptable risk level.
The acceptable risk level is defined by the expected utility which is affected by the long-term experience.
If the perceived risk is lower than the acceptable risk level,
then the driver's driving behavior will become riskier.
The driver's behavior will become more careful in the opposite case.
This way of adjusting behavior based on the perceived risk is called risk compensation behavior.
Note that the perceived risk is a subjective one. 

After the decision making, the driver controls the body to complete the driving operation.
On the side of the vehicle, the vehicle receives the driver's operation and therefore it moves.

\section{Driving behavior model for driving with DAS}

First of all, the process of trust should be discussed in order to prevent over-trust. 
Lee et al. proposed a conceptual model that represents a process of trust in the automation system and describes the dynamics of trust with the role of context~\cite{lee2004trust}.
In this conceptual model, the user gets information from the automation system and forms belief at first.
Then, trust in the automation system is further developed through belief.
The user decides the intention by combining some factors such as self-confidence and perceived risks.
Finally, the reliance action is outputted based on the intention.
Lee's model provides a basic framework of trust.
In this paper, Lee's model has been further refined for our purposes.

Based on the driving behavior model for manual driving, this paper proposes an extended model to describe the driving behavior with the DAS.
This model not only takes mental model and risk compensation behavior of the driver into consider but also considers the trust of the driver in the DAS.
Figure ~\ref{fig:BL2} shows the driving behavior model proposed in this study.

\subsection{Driving automation system}
The DAS is equipped with different kinds of sensors to measure information about driver's operation, vehicle's states, and driving environment.
For example, the driver's operation can be represented by the steering wheel angle, accelerator opening rate, and brake master-cylinder pressure~\cite{Liu2017visualization}.
These sensor information can be measured by the in-vehicle sensors via a controller area network~(CAN).
In addition, cameras and LiDARs can be used to measure information about the driving environment.
The hazards in the driving environment, e.g. 
pedestrians, obstacles, can be recognized by using machine learning methods~\cite{sun2006road,weber2016deeptlr,schneider2016semantic}.
Based on the vehicle state and the result of environment recognition, 
the DAS sends signals to the control system to achieve driving tasks.
Moreover, the DAS presents information about system states or operation instructions to the driver via HMI.
Note that it may trigger over-trust in the DAS if the HMI of the DAS is not designed properly.

\begin{figure*}[tb]
\begin{center}
\includegraphics[width=1\linewidth]{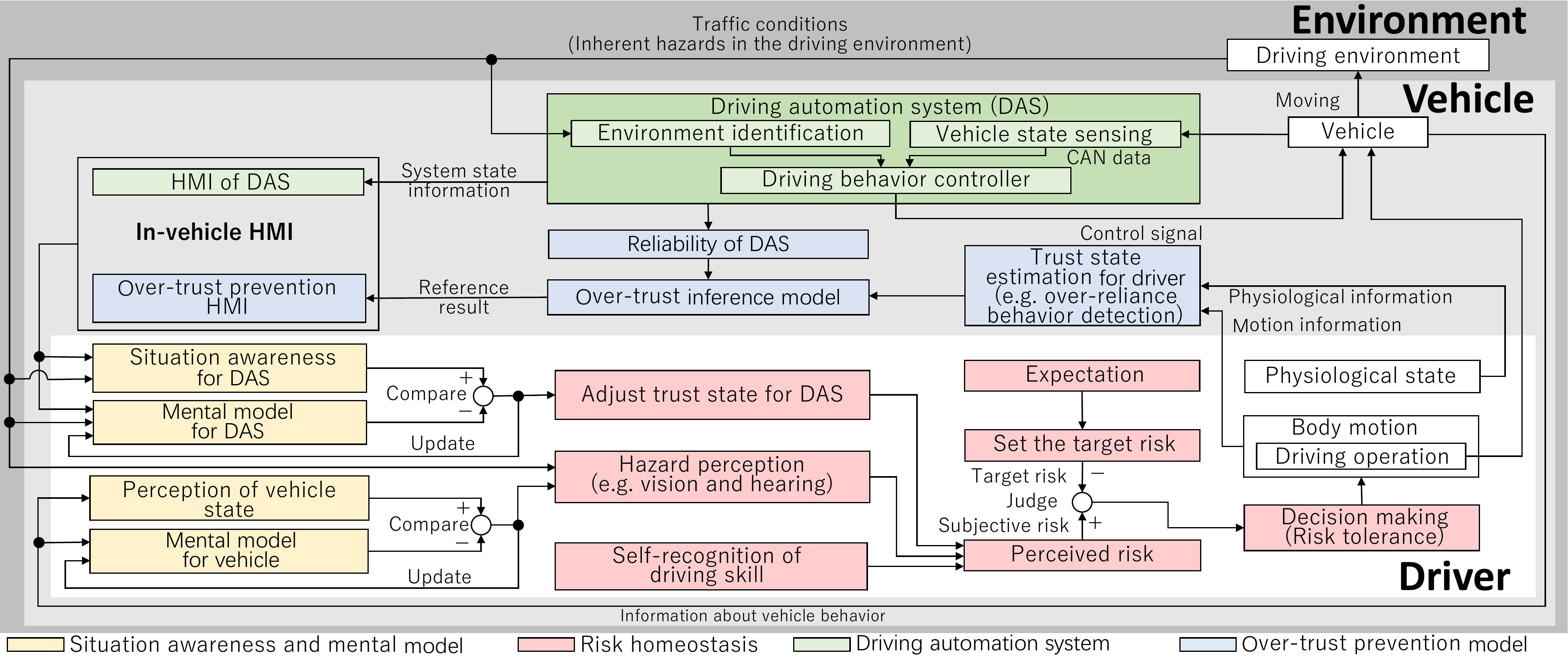}
\end{center}
\captionsetup{justification=centering}
\caption{Driving behavior model considering the mental model, the driver's trust in DAS, and the risk compensation behavior.
(Case 3: when driving vehicle with DAS which equips an over-trust prevention HMI)
}
\label{fig:BL3}
\end{figure*}

\subsection{Trust generation based on a mental model of DAS}
Similar to the generation of the mental model of the vehicle explained before,
a mental model of the DAS also can be generated through the experience of the driver who drives the vehicle with the DAS many times.
This mental model of the DAS represents the driver's understanding of the DAS, such as the foundation, the purpose, the process, and the performance of the DAS.
Consequently, the driver will adjust the trust state in the DAS based on the generated mental model of the DAS.
The driver recognizes the situation of the DAS by the feedback information from the HMI of the DAS.
Then the driver compares the predicted result by using the mental model with the actual situation of the DAS in order to adjust his/her trust state.

\subsection{Risk homeostasis for driving with DAS}
The driver will perceive the risk from the hazards in driving environment and the self-recognition of driving skill when the driver manually drives the vehicle.
Moreover, the driver drives the vehicle with the DAS, the trust in the DAS also affects the risk perception.
As a simple example, the driver predicts a high risk when trust in the DAS is reduced.
After perceiving the risk, the driver makes the decision of driving operation based on the risk compensation behavior the same as the case of manually driving.

Finally, the vehicle is eventually controlled by combining the operations of the driver and the control signals from the DAS.
After the vehicle moving, the driver will recognize the new situation of the vehicle, the DAS and the driving environment, through the movement of the vehicle.
These processes will be looped during driving.

\section{Mechanisms of over-trust and over-trust prevention process}
In order to prevent the over-trust in the DAS, an over-trust prevention process must be considered based on the driving behavior model for driving with the DAS.
This process includes two parts: 1) over-trust inference model, and 2) an over-trust prevention HMI. 
It is shown in Fig.~\ref{fig:BL3}.

\subsection{Over-trust inference model}
The over-trust inference model is proposed based on the assumed mechanisms of over-trust, which will be elaborated next.
According to the two judgment conditions of over-trust in the DAS:
1) the driver is trusting in the DAS,
and 2) the DAS cannot respond to driving tasks,
the over-trust in the DAS can be inferred in the driving process.
In order to detect the driver's over-trust in the DAS, it is necessary to estimate the driver's trust states.

For the first condition, the driver's physiological information, the motion information, and the driving operation are expected to be used to estimate the driver's trust in the DAS.
For example, the motion of the driver can be measured by using a motion capture system~\cite{yamada2016high,cao2017realtime}.
Based on the measured motion information, the operating frequency and the operating quantity may be used to represent the driving involvement.
If the driving involvement of the driver is low, then it means that the driver hands over the driving task to the DAS.
It also means that the driver is trusting in the DAS.
Meanwhile, a face detecting system~\cite{bergasa2006real,ji2004real} can be used to measure the driver's face direction during driving with the DAS.
If the face stays downward for a long time, it represents that the driver may be seeing the mobile phone or reading a book or something.
It also means that the driver is trusting in the DAS.
This paper does not propose a specific mathematical model to estimate the driver's trust in the DAS, but a machine learning-based monitoring system~\cite{simic2016driver} is highly expected.

For the second condition, an evaluation system can be expected to estimate the ability of the DAS on real-time.
For example, the defects detecting system may be used to detect the generated control signal of the DAS is different from the normal driving behaviors~\cite{liu2018defect}. 
Meanwhile, the reliability of the DAS should be evaluated, too~\cite{akai2018simultaneous}.

In summary, the driver's over-trust in the DAS will be inferred by using the estimated trust states of the driver and the evaluated reliability of the DAS.
The mathematical model of inferring the driver's over-trust is a great challenge for this study.

\subsection{Over-trust prevention HMI}
If the predicted ability of the DAS becomes too low to respond to the situation, and the machine learning model estimates that the driver still trusts in the DAS, then an over-trust prevention HMI will be activated. 
It will give an alert and some useful information about the ability of the DAS.
Moreover, the DAS will control the vehicle to perform a minimal risk maneuver, e.g. slow down and pull over, when a driver does not respond to the alert in a period of time.

\section{Conclusion}
The final goal of this study is preventing the driver's over-trust in the DAS, especially for the levels one to three of DAS.
As a first step, this paper discussed the definition and properties of over-trust in the DAS.
Therefore, a mechanism of over-trust and the driving behavior model were proposed. 

As a future work, there are two tasks to be solved.
\begin{description}
\item[1) Estimating the driver's trust states in DAS]~\\ 
In the previous studies~\cite{abe2018driver,desai2013impact}, the trust is evaluated by using the psychological scale.
There is a study~\cite{itoh2012toward} which estimates the driver's trust states by using the physiological index and the driving operators.
Therefore, we will try to use a machine learning approach to estimate the driver's trust states by using the physiological index, driving operators, and the motions of the driver during driving the vehicle with the DAS.

\item[2) Developing an over-trust prevention HMI]~\\
It is necessary for the driver to correctly recognize the situation of the DAS in order to make the driver trust the DAS appropriately.
Consequently, we have to propose and verify the HMI which provides effective information to help the driver to do:
1) perception of the elements in the environment, 2) comprehension of the current situation, and 3) projection of future state~\cite{endsley1995toward}.
In the HMI, we have to consider the problems about what kind of information should be provided, when to provide information, and how to provide information. 
\end{description}

\bibliographystyle{ieeetr}	
\bibliography{sample-base}

\end{document}